\documentclass[12pt]{article}
\usepackage{graphicx,amsfonts,amsmath,amssymb,mathrsfs, comment}
\usepackage{citesort}
\usepackage{url}

\newcommand{\db}{de$\,$Broglie}
\newcommand{\Bm}{Bohmian mechanics}

\def\pa{\partial}

\begin{document}
\vspace*{1.0cm}
\noindent
{\bf
{\large
\begin{center}
On the Uniqueness of Quantum Equilibrium in Bohmian Mechanics 
\end{center}
}
}

\vspace*{.5cm}
\begin{center}
S.\ Goldstein \\
Departments of Mathematics and Physics\\
Rutgers, The State University of New Jersey \\
Hill Center \\
110 Frelinghuysen Road, Piscataway, NJ 08854-8019, USA \\
E--mail: oldstein@math.rutgers.edu
\end{center}

\begin{center}
W.\ Struyve \\
Perimeter Institute for Theoretical Physics \\
31 Caroline Street North, Waterloo, Ontario N2L 2Y5, Canada \\
E--mail: wstruyve@perimeterinstitute.ca
\end{center}

\begin{abstract}
\noindent
In \Bm\ the distribution $|\psi|^2$ is regarded as the equilibrium
distribution. We consider its uniqueness, finding that it is the unique
equivariant distribution that is also a local functional of the wave function
$\psi$.
\end{abstract}

\section{Introduction}
\Bm\ (often called the \db-Bohm theory) yields the same predictions as
standard quantum theory provided the configuration of a system with wave
function $\psi$ is random, with distribution given by $|\psi|^2$. This
distribution, the {\em quantum equilibrium distribution}
\cite{valentini91,durr92},  satisfies the
following natural property: If the distribution of the configuration at
some time $t_0$ is given by $|\psi_{t_0}|^2$, then the distribution of the
configuration at any other time $t$ will be given by $|\psi_t|^2$---i.e.,
with respect to the wave function it will have the same functional form at
the other time---provided, of course, that the wave function evolves
according to Schr\"odinger's equation between the two times and the
configuration evolves according to the law of motion for Bohmian mechanics.
This property was already emphasized by \db\ in 1927 \cite{debroglie28} and was later
formalized and called {\em equivariance} by D\"urr {\em et al.}\
\cite{durr92}, who used it to establish the typicality of empirical
statistics given by the quantum equilibrium distribution.

The notion of equivariance is a natural generalization of that of the
stationarity of a distribution in statistical mechanics and dynamical
systems theory \cite{durr92}. Just as stationarity is regarded as a basic
requirement for a description of equilibrium in statistical mechanics, one
can regard equivariance as a basic requirement for what might be called
equilibrium in \Bm. Of course, this equilibrium need not be a complete
equilibrium, since the wave function in general changes with time and need
not be in equilibrium---even if the configuration is. Rather, equivariance
concerns an equilibrium relative to the wave function: a quantum
equilibrium.

An interesting question which then arises is whether the quantum
equilibrium distribution $|\psi|^2$ is the unique equivariant distribution. In this paper we show that
$|\psi|^2$ is the only local functional of the wave function that is
equivariant.

The uniqueness proof is of particular value for the approach of D\"urr {\em
  et al.}\ \cite{durr92,durr93,durr96,maudlin07} to explaining equilibrium
in Bohmian mechanics, an approach first advocated by Bell \cite{bell81}.
D\"urr {\em et al.}\ base their justification of the $|\psi|^2$
distribution on a ``typicality'' argument. They argue that a ``typical''
Bohmian universe yields $|\psi|^2$ probabilities as empirical
distributions. What this means is that the set of initial configurations of
the universe that yield the $|\psi|^2$ distribution is very large: it has
measure near one for the measure $P_e^\Psi$ having density $|\Psi|^2$, with
$\Psi$ the wave function of the universe. One reason $P_e^\Psi$ is invoked
is that it is equivariant.

After recalling Bohmian mechanics in Section \ref{\Bm}, we define in
Section \ref{Equivariance} the notion of equivariance, providing some
illustrative examples. Some of these touch upon the connection between  the
uniqueness of equivariant distributions for Bohmian mechanics and the
notion of the ergodicity of  a dynamical system, a connection that is
developed in Sections \ref{stat} and \ref{erg}. While some familiarity with
elementary ergodic theory would be helpful for some of the discussion in
Section \ref{Equivariance}, the uniqueness results for the quantum equilibrium
distribution presented in Sections \ref{u} and \ref{?} require no such
familiarity.

\section{\Bm}\label{\Bm}
In  \Bm\ the state of a quantum system is given by the positions of its particles as
well as its wave function; the motion of the particles is
determined by the wave function. For a system of $N$ spinless particles the
wave function $\psi_t(q)=\psi_t(q_1, \dots, q_{3N})$ is a complex-valued
function on the 
configuration space $\mathbb{R}^{3N}$, and  satisfies the
non-relativistic Schr\"odinger equation
\begin{equation}
i\hbar \partial_t \psi_t(q) =H\psi_t(q)= \left(- \sum^M_{k=1}  \frac{\hbar^2}{2m_k} \partial^2_{q_k} + V(q)   \right)  \psi_t(q)\,,
\label{1}
\end{equation}
with $M=3N$, $\partial_{q_k}=\partial /\partial {q_k}$ and where
$m_1=m_2=m_3$ is the mass of the first particle and similarly for the other
particles. The particles move in physical space $\mathbb{R}^{3}$. We denote
the actual positions of the particles by $\mathbf Q_i\in
\mathbb{R}^{3}$. Thus the actual configuration $Q$ of the system of
particles, collectively representing their $N$ actual positions, is given
by the vector $Q=(Q_1, \dots, Q_{M})=(\mathbf Q_1, \dots, \mathbf Q_N) \in
\mathbb{R}^{M}={(\mathbb{R}^{3})}^N$. (The Cartesian coordinates of the
first particle are given by $\mathbf Q_1=(Q_1,Q_2,Q_3)$ and similarly for
the other particles.) The possible trajectories $Q_t$ for the system of
particles are given by solutions to the {\em guidance equation}
\begin{equation}
\frac{dQ_{t}}{dt} = v^{\psi_t}(Q_t) \,,
\label{2}
\end{equation}
where the velocity field $v^{\psi}=(v^{\psi}_1,\dots,v^{\psi}_M)$ on $\mathbb{R}^{M}$ is given by
\begin{equation}
v^{\psi}_k(q) = \frac{\hbar}{m_k}  {\textrm{Im}} \frac{\partial_{q_k} \psi(q)}{\psi(q)} \,.
\label{2.0001}
\end{equation}

We denote the flow associated to the velocity field by $q_t:
\mathbb{R}^{M}\to \mathbb{R}^{M}$.\footnote{The Bohmian dynamics, defined
by equations (1--3), is well defined on the subset of
$L^2(\mathbb{R}^{M})\times \mathbb{R}^{M}$ consisting of pairs $(\psi,q)$
with $\psi$ sufficiently smooth and $q$ such that $\psi(q)\neq0$, see
\cite{berndl96}. We shall usually ignore such details.} Thus $Q_t=q_t(q)$
is the solution to the guidance equation for which $Q_0=q$, so that\
$q_{0}(q)=q$. In this notation we have suppressed the dependence on the
wave function. We keep the initial time $t=0$ fixed throughout the paper,
and let $\psi$ usually denote the initial wave function, so that $\psi_0 =
\psi$.

\section{Equivariance}\label{Equivariance}
Suppose we have a (measure-valued) functional $P:\psi \mapsto P^{\psi}$ from (nontrivial,
i.e.\ not everywhere $0$) wave functions to probability distributions on
configuration space ${\mathbb{R}^M}$.  There exist then two natural time
evolutions for $P^{\psi}$. On the one hand, with $\psi_t(q)=
e^{-iHt/\hbar}\psi(q)$ a solution to the Schr\"odinger equation with
initial wave function $\psi_{0}(q)=\psi(q)$, we have the probability
distribution $P^{\psi_t}$ for all $t \in \mathbb{R}$. On the other hand,
under the Bohm flow (\ref{2},\,\ref{2.0001}) the distribution
$P^{\psi}$ is carried to the distribution $P^{\psi}_t=P^{\psi} \circ q^{-1}_t$ at time $t$. This
means that if the initial configuration $Q_0$ is random, with distribution
$P^{\psi}$, then the distribution of the configuration $Q_t=q_t(Q_0)$ at
time $t$ is $P^{\psi}_t$.

The functional $P$ is called {\em equivariant} \cite{durr92} if
\begin{equation}
P^{\psi}_t = P^{\psi_t} \,, \qquad   \text{for all } t \in \mathbb{R} \,. 
\label{2.005}
\end{equation}
In other words $P$ is equivariant if $P^\psi$ retains its form as a
functional of the wave function $\psi$ when the time evolution of the
distribution is governed by the flow $q_t$ associated to the velocity field
$v^{\psi_t}$. When the equivariant functional $P$ is given by a density,
i.e., when it is of the form $P^{\psi}(dq)= p^{\psi}(q)dq$, we will also
call the density-valued functional $p:\psi \mapsto p^{\psi}(q)$ equivariant. This will of
course be so precisely when $p^{\psi}_t(q) = p^{\psi_t}(q)$ for all $t$,
with $p^{\psi}_t(q)=p^{\psi} (q^{-1}_t(q)) \left| \frac{\partial
q_t}{\partial q}(q^{-1}_t(q))\right|^{-1}$ the density for
$P^{\psi}_t$. (We will also say that the distribution $P^{\psi}$ and the
density $p^{\psi}$ are equivariant when the functionals are.)

We can also characterize equivariance as follows. Suppose $P^\psi$ is given by
the density $p^\psi$. Then the density $p(q,t)=p^{\psi}_t(q)$ satisfies the continuity equation 
\begin{equation}
\partial_t p(q,t) +\sum^M_{k=1}\partial_{q_k} \left(v^{\psi_t}_k(q) p(q,t)  \right)  = 0 \,.
\label{2.002}
\end{equation}
Thus the functional $P$ is equivariant precisely if $\tilde
p(q,t)=p^{\psi_t}(q)$ also satisfies the continuity equation (\ref{2.002})
for all $\psi$. This follows from the uniqueness of solutions of partial
differential equations and the fact that the functions $p^{\psi_t}(q)$ and
$p^{\psi}_t(q)$ are equal at $t=0$.

Let us now give some examples. The first example is the distribution $|\psi|^2$. In the following we don't assume the wave functions to be normalized. If the distributions are given by $|\psi|^2$, then it is natural to normalize the wave functions so that they have $L^2$-norm one. But for  other distributions, other normalizations might be more appropriate.

\vspace{0.5cm}

\noindent
{\bf Example 1} The {\it quantum equilibrium} functional is
$P_e(dq)=p_e(q)\,dq$ where $p_e:\psi \mapsto p^\psi_e = N_e^\psi |\psi|^2$, with
$N_e^\psi=1/\int_{\mathbb{R}^M}|\psi|^2 dq$. Obviously $p_e$, respectively
$P_e$, maps wave functions to probability densities, respectively
probability distributions. This functional is equivariant since
$p^{\psi_t}_e$ satisfies the continuity equation ({\ref{2.002}}) for all
wave functions $\psi$.
\vspace{0.3cm}

In general, whether or not a distribution $P^\psi$ is equivariant would be
expected to depend on the potential $V$. Note, however, that the quantum
equilibrium distribution $P_e$ is equivariant for all $V$.

\vspace{0.5cm}

\noindent
{\bf Example 2} Suppose $\phi$ is a real-valued eigenstate of the
Hamiltonian $H$, for example the ground state. For this stationary state the
associated velocity field $v^{\phi_t}$ (\ref{2.0001}) vanishes, so that the
Bohm motion is trivial in this case. Thus any functional $P$ will trivially
obey (\ref{2.005}) for $\psi=\phi$.

\vspace{0.3cm}

The previous example illustrates the fact that equivariance is a property
of a mapping $\psi\mapsto P^{\psi}$; it concerns a family $\{P^{\psi}\}$ and
not merely the satisfaction of (\ref{2.005}) for a single wave function
$\psi$.  Equivariance means that $P^{\psi_t}=P^{\psi}_t$ for {\em
all} wave functions $\psi$ in the Hilbert space. 

We may also consider the equivariance of a functional $P$ defined on an
invariant subset of Hilbert space: Let $\mathscr{I}$ be an invariant set of
wave functions, i.e., such that $\psi \in \mathscr{I}$ if and only if
$\psi_t=e^{-iHt/\hbar}\psi \in \mathscr{I}$. We say that the functional
$\psi \mapsto P^\psi$, defined for $\psi \in \mathscr{I}$, is {\it
equivariant on} $\mathscr{I}$ if (\ref{2.005}) is obeyed by all $\psi \in
\mathscr{I}$.

We have so far not explicitly imposed any conditions on the
distribution-valued functional $P^{\psi}$ beyond equivariance. A condition
that would be natural is that the functional be projective,
i.e., that if $\psi'$ is a (non-vanishing) scalar multiple of $\psi'$ then
$P^{\psi'}=P^{\psi}$, but we shall not do so. We shall, however, insist on
the following: When we speak of an equivariant functional $P^\psi$, it is
to be understood that the mapping $\psi\mapsto P^\psi$ is measurable. When
$P^{\psi}$ is given by the density $p^{\psi}$, the measurability of
$P^{\psi}$ amounts to that of $p^{\psi}(q)$ as a function of $\psi$ and
$q$. Measurability is the weakest sort of regularity condition invoked in
analysis, probability theory, and ergodic theory, much weaker than
differentiability or continuity.  We do not wish to specify here precisely
what is meant by the measurability of $P^{\psi}$ (or of $p^{\psi}(q)$),
since the main result of this paper involves a much stronger condition,
that $P^{\psi}$ be suitably local. As a rule of thumb, however, we can say
the following: Any mapping $\psi\mapsto P^\psi$ given by an explicit formula
will be measurable. 

In order to appreciate the importance of measurability, one should note
that when a dynamical system is analyzed, it is often necessary to consider
random initial conditions. For the Bohmian system the initial condition is
given  by the  quantum state $\psi$ as well as  the initial configuration,
and hence one should allow for the possibility that the initial wave
function $\psi$ is random, with distribution $\mu(d\psi)$. When this is
combined with a functional $P^{\psi}(dq)$, one is naturally led to consider
the joint distribution $\mu(d\psi)P^{\psi}(dq)$ of $\psi$ and $q$, see
Section \ref{stat}. But this will be meaningful---i.e., define a genuine
probability distribution---only when $P^{\psi}$ is measurable.  

Furthermore, there is a sense in which the equivariance condition
(\ref{2.005}) says that $P^\psi$ is a constant of the motion for the
Schr\"odinger evolution of wave functions: With each
$\psi\in\mathscr{H}=L^2(\mathbb{R}^{M})$ associate a ``fiber''
$\Gamma_{\psi}$, namely the set of probability distributions on
configuration space $\mathbb{R}^{M}$. The Bohm flow acting on distributions
provides a natural identification of $\Gamma_{\psi_t}$ with $\Gamma_{\psi}$
(and in fact defines a connection on the fiber bundle
$\mathscr{H}\times\Gamma = \{\,(\psi,\mu)\,|\,\psi\in\mathscr{H},
\mu\in\Gamma_{\psi}\}$). The equivariance condition (\ref{2.005}) then says
that the function $P^{\psi}$ is a constant of the Schr\"odinger motion
under this identification. 

Now if a dynamical system is ergodic, there can be no nontrivial functions
(i.e., functions that are not almost everywhere equal to a constant) that
are constants of the motion. However, it is understood that only measurable
functions are to be considered; in fact, there are more or less always many
nontrivial constants of the motion that are not measurable. Any function of
the orbits of the motion will define a constant of the motion. Most such
constants of the motion will be nontrivial, and these will also fail to be
measurable when the dynamics is ergodic.

Similarly, one might expect that there will more or less always be a great
many functionals satisfying (\ref{2.005}) if measurability is not demanded,
and this is indeed the case, as we indicate in the next example. (See
Section \ref{erg} for more on equivariance and ergodicity.)

\vspace{0.5cm}

\noindent
{\bf Example 3} For any fixed $\psi$ let ${\cal
O_\psi}=\{e^{-iHt/\hbar}\psi\}\equiv\{\psi_t\}$ denote the orbit of $\psi$
under the Schr\"odinger evolution---the smallest invariant set containing
$\psi$. If ${\cal O_\psi}$ is not a periodic orbit (one such that $\psi_t =
\psi$ for some $t\neq0$), we may let $P^\psi$, for this $\psi$, be {\it
any} probability distribution on configuration space, and extend it to
${\cal O_\psi}$ via (\ref{2.005}). The resulting function $P$ is then
obviously equivariant on ${\cal O_\psi}$. If ${\cal O_\psi}$ is periodic,
let $P=P_e$ on ${\cal O_\psi}$. In this way we may obtain a great many
different functionals $P$---one for each assignment of probability
distributions to representatives of each non-periodic orbit---defined on the
union of all orbits, and hence for all $\psi$ in Hilbert space. All of them
obey (\ref{2.005}) for all $\psi$. Most of these, however, will not be
measurable, and hence should not count as equivariant functionals.  

\vspace{0.3cm} 

In the previous example, suppose we were to choose $P^\psi$ in an explicit
way, for example as in Example 2, on the representatives. It might seem
then, on the one hand, that we have provided in effect an explicit formula
for the functional $P$ constructed in this way, so that it would then be
measurable. On the other hand, if the Bohmian dynamics is suitably ergodic,
see Section \ref{erg}, as is likely often to be the case, $P$ (if it is
given by a density) must then agree with $P_e$ on many non-periodic orbits,
which it clearly does not. What gives? The answer is that the specification
just mentioned is much less explicit than it might at first appear to be,
since in general there is no canonical way to choose a representative for
each orbit, and the functional so constructed need not be measurable.

A flow on the line or an autonomous flow on the plane can't have strong
ergodic properties. One might thus expect the Bohm motion on the line to
also fail to have strong ergodic properties. That this is so was shown in
\cite{goldstein99}. Accordingly, since dynamical systems that are not
ergodic have many stationary distributions, one should expect there to be a
great many distributions that are
equivariant for this case.

\vspace{0.5cm}

\noindent
{\bf Example 4} Consider a Bohmian particle moving on the line. Since
trajectories can't cross, it is easy to see that the function $F(\psi, q)=
P_e^\psi((-\infty, q))$ is a constant of the motion, $F(\psi_t,
q_t(q))=F(\psi, q)$. For fixed $\psi$, $F$ is a map $\mathbb R\to [0,1]$,
and for every probability distribution $\mu$ on $(0,1)$ there is an
equivariant functional
\begin{equation}
P_{\mu}^{\psi}(B)=\mu(F(\psi,B))\,,
\end{equation}
the image of $\mu$ under $F_\psi^{-1}$, the inverse of the map $q\mapsto
F(\psi,q)$. (When $\mu$ is the Lebesgue measure, $\mu(dq)=dq$, we have that
$P_\mu=P_e$.)  Perhaps the simplest way to understand this is in terms of
the change of variables $(\psi,q)\mapsto(\psi,\tilde q)$ with $\tilde
q=F(\psi, q)$. In these new coordinates the Bohmian dynamics becomes
trivial: $\psi$ evolves as usual according to Schr\"odinger's equation and
$\tilde q$ does not change under the dynamics. Thus any distribution $\mu$
for $\tilde q$ defines an equivariant functional.\footnote{Moreover, every
equivariant functional for a particle on the line corresponds a.e.\ to a
(possibly different) choice $\mu$ for each ergodic component of the
Schr\"odinger dynamics.}

\vspace{0.3cm}

A difference between the functional in Example 1 and those in (Example 3
and) Example 4 is that the former is a local functional, whereas the latter
(except for the quantum equilibrium functional) are not. We call a functional $p^{\psi}$ {\em local} if $p^{\psi}(q)$ can
be written, up to normalization, as a (sufficiently differentiable)
function of $q$, $\psi(q)$, and finitely many derivatives of $\psi$,
evaluated at $q$. That is, for a local functional $p^{\psi}$ we can write
\begin{equation}
p^{\psi}(q)=N^{\psi}g^{\psi}(q)
\label{3.001}
\end{equation} 
where $N^{\psi}$ does not depend on $q$ and where
\begin{equation}
g^{\psi}(q) =g\left(q,\psi(q),\dots, \partial^{n_1}_{q_1} \dots  \partial^{n_M}_{q_M}\psi(q),\dots\right) 
\label{sl}
\end{equation} 
depends on at most finitely many partial derivatives of $\psi$ (and is
sufficiently differentiable). We shall say that a functional of the form
(\ref{sl}) is {\em strictly local}. (A local density $p^{\psi}(q)$, because
of the normalization factor $N^{\psi}$, need not be strictly local.) We
note that a local functional that, as demanded above, is differentiable
will of course be measurable. In fact, for measurability,
continuity---indeed mere measurability of $g$---would suffice.

In the following section we will see that equivariance, together with the
requirement that the functional be local, leads uniquely to quantum
equilibrium $p_e$.

\section{Uniqueness of equivariant densities}\label{u}
Let $p:\psi \mapsto p^{\psi}$ be a functional from wave functions to probability densities. We show that $p$ is uniquely given by $p_e$, with $p^{\psi}_e=N^\psi_e|\psi|^2$ as in Example 1, under the assumptions that $p^{\psi}$ is equivariant and local. The locality implies that $p^{\psi}$ can be written in the form $p^{\psi}(q)=N^{\psi}_g g^{\psi}(q)$, where $N^{\psi}_g = 1/\int_{\mathbb{R}^M} g^{\psi}(q) dq$ and where $g^{\psi}(q)$ is a strictly local functional, see (\ref{sl}). We split the proof into two parts, successively showing that: 
\begin{itemize}
\item[(P1)]
$g^{\psi_t}(q)$ satisfies the equation
\begin{equation}
\partial_t g^{\psi_t}(q) + \sum^M_{k=1}\partial_{q_k} \left(v^{\psi_t}_k(q)  g^{\psi_t}(q) \right)   + hg^{\psi_t}(q)  = 0 \,,
\label{3.00202}
\end{equation} 
with $h$ a constant, i.e., independent of $q$ and the wave function.
\item[(P2)]
$p^{\psi}(q) = p^{\psi}_e(q)=N^\psi_e |\psi(q)|^2$.
\end{itemize}

We now give the proofs.\\

\vspace{-0.2cm}
\noindent
{\em Proof of} (P1): Equivariance implies that $p^{\psi_t}(q)$ satisfies the continuity equation (\ref{2.002}). Since $p^{\psi_t}(q)=N^{\psi_t}_g g^{\psi_t}(q)$ the continuity equation for $p^{\psi_t}(q)$ can be written as
\begin{equation}
\frac{1}{g^{\psi_t}(q)}\left( \partial_t g^{\psi_t}(q) + \sum^M_{k=1}\partial_{q_k} \left(v^{\psi_t}_k(q)  g^{\psi_t}(q) \right)   \right) =  -\partial_t \ln N^{\psi_t}_g 
\label{3.002}
\end{equation} 
(wherever  $g^{\psi_t}(q)>0$). 

Let us introduce the functional $h:\psi \mapsto h^\psi$, from wave functions to the real numbers, defined by
\begin{equation}
h^{\psi_t}  =  \partial_t \ln N^{\psi_t}_g\,. 
\label{3.002001}
\end{equation} 
Since $\partial_t \ln N_g^{\psi_t}$ is independent of $q$, $h^{\psi_t}$ is well-defined as a real number. We will show that this functional is constant, i.e.\ independent of $\psi$. 

First note that $\partial_t g^{\psi_t}(q)$ can be expressed as a function
of $q$ and of the variables $\partial^{m_1}_{q_1} \dots
\partial^{m_M}_{q_M}\psi_t(q)$. This is because $g^{\psi}$ is a strictly local functional and because the time derivatives of any of the variables $\partial^{m_1}_{q_1} \dots  \partial^{m_M}_{q_M}\psi_t(q)$ can be replaced by spatial derivatives by making use of the Schr\"odinger equation. As a result we have from (\ref{3.002}) that
\begin{equation}
h^{\psi} = h \left(q,\psi(q),\dots, \partial^{n_1}_{q_1} \dots  \partial^{n_M}_{q_M}\psi(q),\dots \right) \,, 
\label{3.00201}
\end{equation} 
so that $h^{\psi}$ is a strictly local functional. 

It follows that $h^{\psi}=h^{\psi'}$ for any two wave functions $\psi$ and
$\psi'$ for which all derivatives agree at a configuration $q \in
{\mathbb{R}^M}$. But this means that for any $\psi$ and $\psi'$,
$h^{\psi}=h^{\psi'}$, since there is always a third wave function $\psi''$
such that all the derivatives of $\psi$ and $\psi''$ agree at one
configuration $q \in {\mathbb{R}^M}$ and such that all the derivatives of
$\psi'$ and $\psi''$ agree at another configuration $q' \in
{\mathbb{R}^M}$. 

Thus $h^\psi$ is independent of $\psi$. We write $h^\psi=h$. The continuity equation (\ref{3.002}) then reduces to (\ref{3.00202}).

\vspace{0.2cm}
\noindent
{\em Proof of} (P2): Let us introduce the functional $f^{\psi}(q) =
g^{\psi}(q)/ |\psi(q)|^2$.\footnote{$f^{\psi}(q)$ is defined on $\{q\in
\mathbb{R}^M\,|\,\psi(q)\neq0\}$. Since the Bohm flow (\ref{2},\,\ref{2.0001}) is defined
only on this set, we consider only densities on this set, i.e., for which
$g^{\psi}>0$ only on this set.} From the continuity equation for $|\psi_t(q)|^2$ and the equation (\ref{3.00202}) for $g^{\psi_t}(q)$ it follows that
\begin{equation}
\frac{df^{\psi_t}}{dt} + h f^{\psi_t} = 0\,,
\label{3.01}
\end{equation}
with
\begin{equation}
\frac{d}{dt} = \pa_t +  \sum^M_{k=1}  v^{\psi_t}_k \partial_{q_k} \,.
\label{3.1}
\end{equation}
Because $f^{\psi}$ is a strictly local functional we have that
\begin{equation}
f^{\psi}(q) = f \left(q,\psi(q),\dots, \partial^{n_1}_{q_1} \dots  \partial^{n_M}_{q_M}\psi(q),\dots \right) \,.
\label{3.00203}
\end{equation} 
Relation ({\ref{3.01}}) can therefore be written as
\begin{eqnarray}
0&=& \frac{df^{\psi_t}}{dt} + h f^{\psi_t}  \nonumber\\
 &=& \sum^M_{k=1}  v^{\psi_t}_k \partial_{q_k}f +  \sum_{m_1,\dots,m_M  }  \Bigg( \frac{d }{dt}\left( \partial^{m_1}_{q_1} \dots  \partial^{m_M}_{q_M}\psi_{t,r} \right)  \frac{ \partial f}{\partial (\partial^{m_1}_{q_1} \dots  \partial^{m_M}_{q_M}\psi_{t,r})} \nonumber\\
&& + \frac{d }{dt}\left( \partial^{m_1}_{q_1} \dots  \partial^{m_M}_{q_M}\psi_{t,i} \right)  \frac{ \partial f}{\partial (\partial^{m_1}_{q_1} \dots  \partial^{m_M}_{q_M}\psi_{t,i})}  \Bigg)  +   h f \,, 
\label{4.1}
\end{eqnarray}
where $\psi_{t,r}$ and $\psi_{t,i}$ are respectively the real part and the imaginary part of $\psi_t$.

This expression can be rewritten by making use of the Schr\"odinger equation ({\ref{1}}), since for every variable $\partial^{m_1}_{q_1} \dots  \partial^{m_M}_{q_M}\psi_{t,r}$ and $\partial^{m_1}_{q_1} \dots  \partial^{m_M}_{q_M}\psi_{t,i}$ we have that   
\begin{eqnarray}
\frac{d }{dt}\left( \partial^{m_1}_{q_1} \dots  \partial^{m_M}_{q_M}\psi_{t,r} \right) &=&  \partial_t \partial^{m_1}_{q_1} \dots  \partial^{m_M}_{q_M}\psi_{t,r} +  \sum^M_{k=1}  v^{\psi_t}_k \partial_{q_k} \partial^{m_1}_{q_1} \dots  \partial^{m_M}_{q_M}\psi_{t,r} \nonumber\\
&=& - \sum^M_{k=1} \frac{\hbar}{2m_k} \partial^2_{q_k} \partial^{m_1}_{q_1} \dots  \partial^{m_M}_{q_M}\psi_{t,i} + \frac{1}{\hbar} \partial^{m_1}_{q_1} \dots  \partial^{m_M}_{q_M} \left( V\psi_{t,i} \right) \nonumber\\
&& \mbox{} + \sum^M_{k=1}  v^{\psi_t}_k \partial_{q_k}\partial^{m_1}_{q_1} \dots  \partial^{m_M}_{q_M}\psi_{t,r}  
\label{4.2}
\end{eqnarray}
and
\begin{eqnarray}
\frac{d }{dt}\left( \partial^{m_1}_{q_1} \dots  \partial^{m_M}_{q_M}\psi_{t,i} \right) &=&  \partial_t \partial^{m_1}_{q_1} \dots  \partial^{m_M}_{q_M}\psi_{t,i} +  \sum^M_{k=1}  v^{\psi_t}_k \partial_{q_k} \partial^{m_1}_{q_1} \dots  \partial^{m_M}_{q_M}\psi_{t,i} \nonumber\\
&=&  \sum^M_{k=1} \frac{\hbar}{2m_k} \partial^2_{q_k} \partial^{m_1}_{q_1} \dots  \partial^{m_M}_{q_M}\psi_{t,r} - \frac{1}{\hbar} \partial^{m_1}_{q_1} \dots  \partial^{m_M}_{q_M} \left( V\psi_{t,r} \right) \nonumber\\
&& \mbox{} + \sum^M_{k=1}  v^{\psi_t}_k \partial_{q_k}\partial^{m_1}_{q_1} \dots  \partial^{m_M}_{q_M}\psi_{t,i} \,. 
\label{4.201}
\end{eqnarray}
In this way the equation (\ref{4.1}) expresses a functional relation between the variables $q$ and all the real variables $\partial^{n_1}_{q_1} \dots  \partial^{n_M}_{q_M}\psi_{t,r}$ and $\partial^{n_1}_{q_1} \dots  \partial^{n_M}_{q_M}\psi_{t,i}$ which has to hold identically, i.e.\ for all possible values of these variables. Since all these variables can be treated as independent, we can show that the function $f$ must be a constant as follows.
       
First select a variable $\partial^{n_1}_{q_1} \dots
\partial^{n_M}_{q_M}\psi_{t,r}$ or $\partial^{n_1}_{q_1} \dots
\partial^{n_M}_{q_M}\psi_{t,i}$ such that $f$ depends on this variable and
such that, if $f$ depends on another variable $\partial^{{\bar n}_1}_{q_1}
\dots \partial^{{\bar n}_M}_{q_M}\psi_{t,r}$ or $\partial^{{\bar
n}_1}_{q_1} \dots \partial^{{\bar n}_M}_{q_M}\psi_{t,i}$, then ${\bar n}_1
\le n_1$. Suppose the selected variable is, say, $\partial^{n_1}_{q_1}
\dots \partial^{n_M}_{q_M}\psi_{t,r}$. Then, from ({\ref{4.1}}),
({\ref{4.2}}) and ({\ref{4.201}}) it follows that the only term in
$df^{\psi_t}/dt + hf^{\psi_t}$ that contains the variable
$\partial^{n_1+2}_{q_1} \dots \partial^{n_M}_{q_M}\psi_{t,i}$ is
\begin{equation}
-\frac{\hbar}{2m_1}  \partial^{n_1+2}_{q_1} \dots  \partial^{n_M}_{q_M}\psi_{t,i}  \frac{ \partial f}{\partial (\partial^{n_1}_{q_1} \dots  \partial^{n_M}_{q_M}\psi_{t,r})}\,.
\label{7}
\end{equation}
Because $\partial^{n_1+2}_{q_1} \dots  \partial^{n_M}_{q_M}\psi_{t,i}$ can be treated as an independent variable, the term above has to be zero. Hence
\begin{equation}
\frac{ \partial f}{\partial \left( \partial^{n_1}_{q_1} \dots  \partial^{n_M}_{q_M}\psi_{t,r}  \right)} = 0\,.
\label{8}
\end{equation}
But this contradicts the fact that $f$ depends on the variable
$\partial^{n_1}_{q_1} \dots \partial^{n_M}_{q_M}\psi_{t,r}$. It follows
that $f$ does not depend on any of the variables $\partial^{m_1}_{q_1}
\dots \partial^{m_M}_{q_M}\psi_{t,r}$ or $\partial^{m_1}_{q_1} \dots
\partial^{m_M}_{q_M}\psi_{t,i}$. Hence we have that $f=f(q)$.

Equation ({\ref{4.1}}) now reduces to 
\begin{equation}
\sum^M_{k=1}  v^{\psi_t}_k \partial_{q_k}f +   h f = 0  
\label{8.1}
\end{equation}
and we can use a reasoning similar to the above to conclude that $\partial_{q_k}f=0$, $k=1,\dots,M$. Hence $f$ is a constant independent of $q$ and the wave function and any of its derivatives. Since $g^\psi(q)=f|\psi|^2$ with $f$ now a constant and since $p^\psi$ is assumed to be a probability density we have that 
\begin{equation}
p^\psi(q)  = p^{\psi}_e(q)=N_e^\psi |\psi(q)|^2 \,.
\label{8.2}
\end{equation}

\section{A stronger result?}\label{?}

There is a weaker version of the locality of the functional
$p^{\psi}(q)=N^{\psi}g^{\psi}(q)$, which we shall call {\em weak locality},
that is worth considering. This requires that $g^{\psi}(q)$ be determined
by $\psi$ in a neighborhood of $q$, i.e., that if $\psi$ and $\psi'$ agree
in some neighborhood of $q$, then $g^{\psi}(q) = g^{\psi'}(q)$. This is
indeed a weaker notion of locality than used earlier, and allows in
particular for $g^{\psi}(q)$ to depend on all derivatives of $\psi$ at $q$.

It is reasonable to ask whether the uniqueness result would continue to be
valid if the equivariant functional $p^\psi$ were assumed only to be weakly
local. We believe that the answer is yes. There is an argument for this
that, while not entirely rigorous, is quite compelling. At the same time,
the argument provides some perspective on our uniqueness result.  It is
this:

The Bohmian dynamics defines a flow on (a subset of) the space
$\mathscr{X}=\mathscr{H}\times\mathbb{R}^{M}$, where $\mathscr{H}=L^2(\mathbb{R}^{M})$
is the Hilbert space of the Bohmian system. We shall denote the action of
this flow by $T_t$, so that for $\eta=(\psi,q)\in\mathscr{X}$, we have that
$T_t\eta=(\psi_t,q_t(q))$. In terms of this flow, the equivariance of the
density $p^{\psi}(q)$ can be conveniently expressed as follows: Let
\begin{equation}\label{G} 
G(\eta)= p^{\psi}(q)/p_e^{\psi}(q)\,.
\end{equation} 
Then the equivariance of $p^{\psi}$
amounts to the requirement that $G$ be a constant of the motion for the
flow $T_t$,
\begin{equation}\label{cm} 
G(T_t\eta)=G(\eta)\,.
\end{equation}
(This is an easy consequence of the fact that $p^{\psi}_e$ is equivariant.)
And uniqueness amounts to the statement that $G$ is constant on (the
relevant subset of) $\mathscr{X}$. This would be so if the flow $T_t$ were
sufficiently ergodic (see Section \ref{erg}): ergodicity means that there
are no nontrivial constants of the motion---that the only constants of the
motion are in fact functions that are almost everywhere constant, and hence
trivially constants of the motion---as would be the case if the set of
possible states $\eta$ consisted of a single trajectory. This, of course,
is impossible. Nonetheless, the ergodicity of a motion on a space means
roughly that the motion is sufficiently complicated to produce trajectories
that almost connect any two points in the space, so that functions that
don't change along a trajectory must be more or less everywhere constant.

In fact, it is easy to see that for uniqueness it is sufficient that $G$ be
constant on the subsets
$\mathscr{X}_\psi=\{(\psi,q)\in\mathscr{X}\,|\,q\in\mathbb{R}^{M}\}$ of
$\mathscr{X}$ corresponding to fixed $\psi$, and for this it is of course
sufficient that $G$ be locally constant on $\mathscr{X}_\psi$, i.e., that
every $q\in\mathbb{R}^{M}$ has a neighborhood $O_q$ such that $G$ is
constant on $\{(\psi,q')\in\mathscr{X}\,|\,q'\in O_q\}$. It is also easy to
see that for uniqueness it is sufficient that
$F(\eta)=g^\psi(q)/|\psi(q)|^2$ be constant on $\mathscr{X}$---or (locally)
constant on $\mathscr{X}_\psi$.

While $F$ is not obviously invariant under the flow $T_t$, it is clearly
{\em quasi-invariant}, which is almost as good: In terms of $F$, (\ref{cm})
becomes
\begin{equation}\label{finv} 
F(T_t\eta)=e^{ht}F(\eta)\,,
\end{equation}
for all $t\in\mathbb{R}$, where $h$ is the constant defined by
(\ref{3.002001}). (That $h^{\psi}$ is constant follows from weak locality
much as it does from locality. Moreover, it seems likely on general grounds
that $h=0$, in which case $F$ would be strictly invariant.)

Now the (weak) locality of $p^{\psi}$ implies that $F$ is invariant under a
much larger set of transformations than the one-dimensional set $\{T_t\}$,
defining an action of the group $\mathbb{R}$ on $\mathscr{X}$. It implies
invariance under the action $T_\phi$ of the infinite-dimensional (additive)
group $\mathscr{N}=\{\phi\in\mathscr{H}\,|\,\phi(q') = 0\text{ in a
neighborhood of $q=0$}\}$, where
$T_\phi\eta=T_\phi(\psi,q)=(\psi+\phi_q,q)$, with
$\phi_q(q')=\phi(q'-q)$. Thus with weak locality we have, in addition to
(\ref{finv}), that for all $\phi\in\mathscr{N}$
\begin{equation}\label{inv} 
F(T_\phi\eta)=F(\eta)\,.
\end{equation}

Now while the action of $\mathbb{R}$ on $\mathscr{X}$ given by the Bohmian
flow $T_t$ may fail to be suitable ergodic, it is hard to imagine this for
the action $T_\xi,\ \xi\in \mathscr{G}$, of the group $\mathscr{G}$
generated by the actions of $\mathbb{R}$ and $\mathscr{N}$ on
$\mathscr{X}$. Indeed, it seems very likely that $\mathscr{X}$ 
consists of a single orbit $\{T_\xi(\psi,q)\,|\,\xi\in \mathscr{G}\}$ of
this action, and more likely still that $\mathscr{G}$  connects any two points
in any sufficiently small neighborhood of any point in $\mathscr{X}_\psi$.

If $h$ were 0 this would imply uniqueness. For general $h$ we have that 
\begin{equation}\label{xinv} 
F(T_\xi\eta)=e^{ht_\xi}F(\eta)
\end{equation}
for all $\xi\in\mathscr{G}$. But what was suggested above for $\mathscr{G}$
should still be true of the subgroup
$\mathscr{G}_0=\{\xi\in\mathscr{G}\,|\, t_\xi=0$\}, under the action of which
$F$ is invariant, and this would imply uniqueness in the general case.

Indeed, consider only the transformations in $\mathscr{G}_0$ of the form
$T_{\phi_2,-t,\phi_1,t}= T_{\phi_2} T_{-t}T_{\phi_1} T_t$, with
$\phi_i\in\mathscr{N}$ and $t\in\mathbb{R}$. Since the dimension of the
set of such transformations should be regarded as roughly twice the
dimension of $\mathscr{X}$,  the set obtained by
applying all such transformations to a given
point  $\eta\in\mathscr{X}$---the range of the mapping
$(\phi_1,\phi_2,t)\mapsto T_{\phi_2,-t,\phi_1,\,t}\eta$---should be all of
$\mathscr{X}$,  at the very least, locally.

The previous argument also suggests that for the uniqueness of the
equivariant distribution, the locality condition can be weakened further to
that of having finite range $r>0$: that $g^\psi(q)$ depend at most on the
restriction of $\psi$ to the ball $B_r$ of radius $r$ centered at $q$. (The
weak locality condition is then that of having finite range $r$ for all
$r>0$.)
 
\section{Equivariance and stationarity}\label{stat}

We have already indicated that an equivariant functional can be regarded as
generalizing the notion of a stationary probability distribution for a
dynamical system---one that is invariant under the time-evolution. We wish
here to tighten this connection a bit, and observe that the equivariance of
the functional $P^\psi$ is more or less equivalent to (it implies and is
almost implied by) the following: For every measure $\mu(d\psi)$ on Hilbert
space $\mathscr{H}$ that is stationary under the Schr\"odinger evolution,
the measure $\mu(d\psi)P^{\psi}(dq)$ is a stationary measure on
$\mathscr{X}=\mathscr{H}\times\mathbb{R}^M$ for the Bohmian dynamics. (The
``almost'' and ``more or less'' refer to the following: The stationarity of
$\mu(d\psi)P^{\psi}(dq)$ implies that the condition (\ref{2.005}) for
equivariance is satisfied by all $\psi$ with the possible exception of a
set of $\psi$'s with $\mu$-measure 0. If there are exceptional $\psi$'s,
$P^{\psi}(dq)$ can be changed, on a set with $\mu$-measure 0 so that it
continues to define the same measure $\mu(d\psi)P^{\psi}(dq)$ on
$\mathscr{X}$, so as to become strictly equivariant.)

A general probability measure on $\mathscr{X}$ can be regarded as of the
form $\mu(d\psi)P^{\psi}(dq)$: $\mu(d\psi)$ is the first marginal, the
distribution of the first component $\psi$ of
$\eta=(\psi,q)\in\mathscr{X}$, and $P^{\psi}(dq)$ is the conditional
distribution of the configuration $q$ given $\psi$, a probability measure on the
fiber of the product space $\mathscr{X}$ that ``lies above $\psi$''.
Consider now any measure on $\mathscr{X}$ of the form
$\mu(d\psi)P^{\psi}(dq)$, with now $\mu$ any measure on $\mathscr{H}$ and
$P^{\psi}(dq)$ a probability measure on $\mathbb{R}^M$. (Here $\mu$ need
not be a probability measure, nor even normalizable.) For this measure to
be stationary $\mu(d\psi)$ obviously must be. Suppose this is so. Then, for
stationarity, we still must have that the measure $P^{\psi}(dq)$ on the
$\psi$-fiber evolves to the correct measure on the $\psi_t$-fiber, namely
$P^{\psi_t}(dq)$ (with the possible exception of a set of $\psi$'s having
$\mu$-measure 0). But equivariance says more or less precisely that this is
so: it says that for all $\psi$, $P^{\psi_t} = {P_t^{\psi}}$, the
measure to which $P^{\psi}$ evolves.

Thus a probability measure on $\mathscr{X}$ is stationary if and only if it
is of the form $\mu(d\psi)P^{\psi}(dq)$ with $\mu$ stationary and
$P^{\psi}$ equivariant.  In particular, the measure
$\mu(d\psi)P_e^{\psi}(dq)$, where $P_e$ is the quantum equilibrium
distribution, is stationary whenever $\mu(d\psi)$ is.  Suppose this is
so. Consider a measure $\mu(d\psi)P^{\psi}(dq)$ having a density with
respect to $\mu(d\psi)P_e^{\psi}(dq)$. This density is given by the
function $G$ (\ref{G}) on $\mathscr{X}$. The measure
$\mu(d\psi)P^{\psi}(dq)$ will be stationary precisely if its density $G$ is
a constant of the motion, consistent with our earlier assertion that this
amounts to the equivariance of $P^{\psi}$.

\section{Uniqueness and ergodicity}\label{erg}

The ergodicity of a dynamical system, defined by a dynamics and a given
stationary probability distribution, is equivalent to the statement that
any stationary probability distribution with a density with respect to the
given one must in fact be the given one.  Thus ergodicity amounts to the
uniqueness, in an appropriate sense, of a stationary measure.  So a
uniqueness statement for an equivariant functional---a uniqueness statement
for quantum equilibrium---can be regarded as expressing a sort of
generalized ergodicity. We wish now to sharpen this connection by observing
that certain uniqueness statements for quantum equilibrium are more or less
equivalent to the ergodicity of certain dynamical systems. (One should bear
in mind that the ergodicity of a dynamical system is usually extremely
difficult to establish.)

The relevant dynamical systems for our purposes here are defined by the
Bohmian dynamics on $\mathscr{X}$, with this space equipped with a
stationary probability measure of the form $\mu(d\psi)P_e^{\psi}(dq)$, with
$\mu(d\psi)$ stationary under the Schr\"odinger dynamics, as described in
Section \ref{stat}. In order for this dynamical system to be ergodic, it
is of course necessary for $\mu(d\psi)$ to be an ergodic measure for the
Schr\"odinger dynamics. Suppose that this is so. Then it is easy to see
that the ergodicity of $\mu(d\psi)P_e^{\psi}(dq)$ under the Bohmian
dynamics amounts to the uniqueness of quantum equilibrium ``modulo
$\mu(d\psi)$'': $\mu(d\psi)P_e^{\psi}(dq)$ is ergodic if and only if every
equivariant density $p^{\psi}$ agrees with quantum equilibrium,
$p^{\psi}=p_e^{\psi}$, for $\mu$-a.e. $\psi$.\footnote{A genuinely
different equivariant distribution $P^\psi$ with density $p^\psi$---one
that does not agree with $P_e^{\psi}$ for $\mu$-a.e. $\psi$---would yield a
stationary probability distribution on $\mathscr{X}$ that is given by a
density with respect to the one arising from $P_e^{\psi}$ but that differs
from it, contradicting ergodicity. Conversely, by the discussion of Section
\ref{stat} and the ergodicity of $\mu$, a stationary probability
distribution on $\mathscr{X}$ that is given by a density with respect to
$\mu(d\psi)P_e^{\psi}(dq)$ must be of the form $\mu(d\psi)P^{\psi}(dq)$
with $P^{\psi}(dq)$ equivariant.}
 
There is, however, perhaps less in this equivalence than first meets the
eye. The set of $\psi$'s of $\mu$ measure 1 for which, as a consequence of
the ergodicity of $\mu(d\psi)P_e^{\psi}(dq)$, we must have that
$p^{\psi}=p_e^{\psi}$ when $p^{\psi}$ is an equivariant density  will be rather
small. The set is large only relative to the ``support'' of $\mu$, an
invariant subset $\mathscr{I_\mu}$ of $\mathscr{H}$, with $\mu$-measure 1,
defined by specified values of the constants of the Schr\"odinger motion
such as $\left<\psi|H^n|\psi\right>,\ n=0,1,2,\dots$. 

For every such ``ergodic component'' $\mathscr{I_\mu}$ of the Schr\"odinger
dynamics, with $\mu(d\psi)P_e^{\psi}(dq)$ also ergodic, we have the
uniqueness of quantum equilibrium for almost all $\psi$ in
$\mathscr{I_\mu}$. Taking the totality of such ergodic components of the
Schr\"odinger dynamics, we obtain the uniqueness of quantum equilibrium for
almost all of the union of these components. In particular, if
$\mathscr{H}$ were completely decomposable into such ergodic components, we
would have the uniqueness of quantum equilibrium for almost all $\psi$ in
$\mathscr{H}$.\footnote{Such a decomposition, of all of $\mathscr{H}$,
probably never exists. For many stationary states $\psi$ the Bohm motion is
trivial, so that, with $\mu$ the uniform distribution on the orbit ${\cal
O}_\psi$ of $\psi$, which is ergodic for the Schr\"odinger dynamics,
$\mu(d\psi)P_e^{\psi}(dq)$ is not ergodic, see Example 2. And for wave
functions belonging to the spectral subspace of $\mathscr{H}$ corresponding
to the continuous spectrum the situation is even worse. For example, for a
free Hamiltonian $H$, with $V=0$, there are no ergodic components to begin
with. There are in fact, in this case, no probability measures on
$\mathscr{H}$ that are stationary under the Schr\"odinger
dynamics. (Consider the free Schr\"odinger dynamics.  As time goes on the
wave function should spread, never to become narrow again. But this
conflicts with Poincar\'e recurrence, and thus implies that there is no
finite invariant measure, and in particular no stationary probability
measure.) And in this case as well, there are, presumably, equivariant
densities $p^{\psi}$ that disagree with $p_e^{\psi}$ for all $\psi$.}

Here is an example of a typical ergodic component of the Schr\"odinger
dynamics, to which the discussion of this section could be applied. Suppose
$\phi_1,\dots,\phi_n$ are eigenstates of the Hamiltonian $H$, with
corresponding eigenvalues $E_1,\dots,E_n$ that are rationally
independent. For $c_j>0,\ j=1,\dots,n$, let
$\mathscr{I}_{c_1,\dots,c_n}=\{\psi\in\mathscr{H}\,|\,\psi=\sum_{j=1}^n c_j
e^{i\theta_j}\phi_j,\ 0\leq \theta_j< 2\pi, \ j=1,\dots,n\}.$ The
Schr\"odinger dynamics on $\mathscr{I}_{c_1,\dots,c_n}$ is quasi-periodic,
with stationary probability distribution, corresponding to a uniform
distribution of the phases $\theta_j$, that is ergodic. 

\section{Properties of quantum equilibrium}

The quantum equilibrium functional $P^\psi=P_e^\psi$ satisfies many natural
conditions, some of which play an important role in the analysis of a
Bohmian universe: 

\begin{itemize}
\item[(i)]  It is {\em universally} equivariant: it is equivariant for
all Schr\"odinger Hamiltonians $H$,  of the form expressed on  the right
hand side  of equation (\ref{1}), i.e., for all $V$ and for all choices
$m_k$ of the masses of the particles.
\item[(ii)] It is projective: $P^{c\psi}=P^\psi$ for every
constant $c\neq0$.
\item[(iii)] It is covariant: $P_R^\psi=P^{R\psi}$ for all the usual
symmetries of non-relativistic quantum mechanics, for example for
space-translations, rotations, time-reversal, Galilean boosts, and
particle permutations. Here $P_R^\psi$ is the distribution to which
$P^\psi$ is carried by the action of $R$ on configurations.
\item[(iv)] It is {\em factorizable}. Suppose a Bohmian system is a
composite of two systems, with Hilbert space
$\mathscr{H}=\mathscr{H}_1\otimes \mathscr{H}_2$ and configuration variable
$q=(q^{(1)},q^{(2)})$.  Then $P^{\psi_1\otimes\psi_2}(dq^{(1)}\times
dq^{(2)})=P^{\psi_1}(dq^{(1)})P^{\psi_2}(dq^{(2)})$. (If $H = H_1\otimes
I_2\, +\, I_1\otimes H_2$, with $I_i$ the identity on $\mathscr{H}_i$, then
it follows immediately from the equivariance of $P$ for the composite
system that the $P^{\psi_i}$ are equivariant for the respective components.)
\item[(v)] More generally, it is {\em hereditary}. Consider a composite
system as in (iv), and suppose that the conditional wave function of, say,
system 1 is $\psi$ when the composite has wave function $\Psi$ and system 2
has configuration $Q^{(2)}$, i.e., that
$\psi(q^{(1)})=\Psi(q^{(1)},Q^{(2)})$. Then the conditional distribution
of the configuration of system 1, given that the configuration of system~2 is
$Q^{(2)}$, depends only on $\psi$ and not on the choice of wave function
$\Psi$ and configuration $Q^{(2)}$ that yields $\psi$: for fixed $\psi$,
$P^\Psi(dq^{(1)}\,|\,Q^{(2)})$ is independent of $\Psi$ and $Q^{(2)}$.
\end{itemize}

It remains to be seen to what extent these properties, individually or in
various combinations,  uniquely characterize quantum equilibrium among
equivariant distributions. (It presumably follows, along the lines of the
discussion in Section \ref{?}, that the satisfaction of the equivariance
condition (\ref{2.005}) for all $V$'s implies uniqueness---with the
exception of the case of a single particle on a line.) Be that as it
may, it is noteworthy that locality alone, with no additional conditions
beyond equivariance, is sufficient to guarantee the uniqueness of quantum
equilibrium.

\section{Acknowledgements}
Discussions with Detlef D\"urr, Michael Kiessling, Owen Maroney, Roderich
Tumulka, Antony Valentini, Hans Westman and Nino Zangh\`i are gratefully
acknowledged. The work of S. Goldstein was supported in part by NSF Grant
DMS--0504504. Research at Perimeter Institute for Theoretical Physics is supported in part by the Government of Canada through NSERC and by the Province of Ontario through MRI.

\end{document}